\documentstyle[12pt]{article}

\newcommand{\be}{\begin{equation}}
\newcommand{\ee}{\end{equation}}
\newcommand{\bea}{\begin{eqnarray}}
\newcommand{\eea}{\end{eqnarray}}

\newtheorem{theorem}{Theorem}
\newtheorem{definition}{Definition}
\newtheorem{proposition}{Proposition}
\newtheorem{corollary}{Corollary}

\begin{document}

\title{{\bf On the differentiability }\\{\bf of Cauchy horizons}\thanks{
1991 Mathematics Subject Classification  53C50, 53C80, 83C75.}
}

\author{Robert J. Budzy\'nski
\\
{\small\em Department of Physics,}
\\
{\small\em Warsaw University,}
\\
{\small\em Ho\.za 69,}
{\small\em 00-681 Warsaw, Poland}
\and
Witold Kondracki and Andrzej Kr\'olak
\\
{\small\em Institute of Mathematics,}
\\
{\small\em Polish Academy of Sciences,}
\\
{\small\em \'Sniadeckich 8,}
{\small\em 00-950 Warsaw, Poland.}
}


\maketitle
\begin{abstract}
Chru\'sciel and Galloway constructed a Cauchy horizon that is
non-differentiable on a dense set.
We prove that in a certain class of Cauchy horizons densely nondifferentiable
Cauchy horizons are generic. We show that our class of densely
nondifferentiable Cauchy horizons implies the existence of densely nondifferentiable 
Cauchy horizons arising from partial Cauchy surfaces
and also the existence of densely nondifferentiable black hole event horizons.
\end{abstract}

\thispagestyle{empty}
\newpage

\setcounter{page}{1}
\section {Introduction}

Recently Chru\'sciel and Galloway \cite{CG98} have constructed an
example of a Cauchy horizon which fails to be differentiable on a
dense subset. In this paper we show that densely nondifferentiable Cauchy
horizons appear to be generic in a certain class of Cauchy horizons.
Chru\'sciel and Galloway have also shown that their example implies
the existence of a densely nondifferentiable black hole event horizon.
They point out that these examples raise definite questions concerning
some major arguments that have been given in the past where smoothness
assumptions were implicitly made.  In the light of these new
examples, it is clear that there is a real need for a deeper
understanding of the differentiability properties of horizons.

In a spacetime with a partial Cauchy surface $S$ the Cauchy horizon
$H(S)$ is the boundary of the set of points where, in theory, one may
calculate everything in terms of the initial data on $S$.  Cauchy
horizons are {\em achronal} (i.e., no two points on the horizon may be
joined by a timelike curve) and this implies that Cauchy horizons
(locally) satisfy a Lipschitz condition. This, in turn, implies that
Cauchy horizons are differentiable almost everywhere.  Because they
are differentiable except for a set of (three-dimensional) measure zero,
it seems that they have often been assumed to be smooth except for a
set which may be more or less neglected.  However, one must remember
in the above that: (1) differentiable only refers to being
differentiable at a single point, and (2) sets of measure zero may be
quite widely distributed.

For $S$ a closed achronal set each point $p$ of a Cauchy horizon
$H^+(S)$ lies on at least one null generator \cite{HE73}. 
However, null generators may or may not remain on the horizon
when they are extended in the future direction.  If a null generator
leaves the horizon, then there is a last point where it remains on the
horizon.  This last point is said to be an {\em endpoint} of the
horizon.  Endpoints where two or more null generators leave the
horizon are points where the horizon must fail to be differentiable
\cite{H92}, \cite{CG98}.  In addition, Chru\'sciel and Galloway
\cite{CG98} have shown that Cauchy horizons are differentiable at
points which are not endpoints.  Beem and Kr\'olak have shown
\cite{BK98} that Cauchy horizons are differentiable at endpoints where
only one generator leaves the horizon.  These results give a complete
classification of (pointwise) differentiability for Cauchy horizons in
terms of null generators and their endpoints. Beem and Kr\'olak have
also shown \cite{BK98} that if we consider an open subset $W$ of the
Cauchy horizon $H^+(S)$ and assume that the horizon has no endpoints
on $W$, then the horizon must be differentiable at each point of $W$
and, in fact, that the horizon must be at least of class $C^1$ on $W$.
Conversely, the differentiability on an open set $W$ implies there are
no endpoints on $W$.

For general spacetimes, horizons may fail to be stable under small
metric perturbations; however, some sufficiency conditions for various
stability questions have been obtained \cite{B95}, \cite{CI96}.

\section{Preliminaries}


\begin{definition}
A space-time $(M,g)$ is a smooth $n-$dimensional, Hausdorff manifold
$M$ with a semi-Riemannian metric $g$ of signature $(-,+,...,+)$, a
countable basis, and a time orientation.
\end{definition}

A set $S$ is said to be {\em achronal} if there are no two points of
$S$ with timelike separation.

We give definitions and state our results in terms of the future
horizon $H^+(S)$, but similar results hold for any past Cauchy horizon
$H^-(S)$.


\begin{definition}
The {\em future Cauchy development} $D^{+}(S)$ consists of all points
$ p \in M$ such that each past endless and past directed causal curve
from $p$ intersects the set $S$. The {\em future Cauchy horizon} is
$H^{+}(S) = \overline{(D^{+} (S))} -I^{-}(D^{+}(S))$.
\end{definition}

Let $p$ be a point of the Cauchy horizon; then there is at least one
null generator of $H^{+}(S)$ containing $p$.  Each null generator is
at least part of a null geodesic of M.  When a null generator of
$H^{+}(S)$ is extended into the past it either has no past endpoint or
has a past endpoint on $edge(S)$ [see \cite{HE73}, p. 203].  However,
if a null generator is extended into the future it may have a last
point on the horizon which then said to be an {\em endpoint} of the
horizon.  We define the {\em multiplicity} [see \cite{BK98}] of a
point $p$ in $H^{+}(S)$ to be the number of null generators containing
$p$.  Points of the horizon which are not endpoints must have
multiplicity one.  The multiplicity of an endpoint may be any positive
integer or infinite.  We call the set of endpoints of multiplicity two
or higher the {\em crease set}, compare \cite{CG98}.  By a basic
Proposition due to Penrose [\cite{HE73}, Prop. 6.3.1] 
$H^+(S)$ is an $n-1$ dimensional
Lipschitz topological submanifold of $M$ and is achronal. Since a
Cauchy horizon is Lipschitz it follows from a theorem of Rademacher
that it is differentiable almost everywhere (i.e.  differentiable
except for a set of $n-1$ dimensional measure zero). This does not
exclude the possibility that the set of non-differentiable points is
a dense subset of the horizon.  An example of such a behaviour was
given by Chru\'sciel and Galloway \cite{CG98}.

Following \cite{BK98} let us introduce the notion of
differentiablility of a Cauchy horizon.  Consider any fixed point $p$
of the Cauchy horizon $H^{+}(S)$ and let $x^{0},x^{1}, x^{2},x^{3}$ be
local coordinates defined on an open set about $p = (p^0, p^1, p^2,
p^3)$. Let $H^+(S)$ be given near $p$ by an equation of the form
$$
x^0 = f_H(x^1, x^2, x^3)
$$
%
The horizon $H^+(S)$ is {\em differentiable} at the point $p$ iff
the function $f_H$ is differentiable at the point $(p^1, p^2, p^3)$.
In particular, if $p = (0, 0, 0, 0)$ corresponds to the origin in the given
local coordinates and if
$$
\Delta x = (x^1, x^2, x^3)
$$
%
represents a small displacement from $p$ in the $x^0 = 0$ plane, then
$H^+(S)$ is differentiable at $p$ iff one has
$$
f_{H}(\Delta x) = f_{H}(0) + \sum a_{i}x^{i} + R_H(\Delta x)
 = 0 + \sum a_{i}x^{i} + R_H(\Delta x)
$$
%
where the ratio $R_H(\Delta x)/| \Delta x |$
converges to zero as $| \Delta x |$ goes to zero.
Here we use
$$
|\Delta x| = \sqrt{ (x^{1})^{2} + (x^{2})^{2} + (x^{3})^{2} }.
$$

If $H^{+}(S)$ is differentiable at the point $p$, then there is a well
defined 3-dimensional linear subspace $N_0$ in the tangent space
$T_{p}(M)$ such that $N_{0}$ is tangent to the 3-dimensional surface
$H^{+}(S)$ at $p$.  In the above notation a basis for $N_{0}$ is given
by $\{a_{i} \partial / \partial x^{0} + \partial / \partial x^{i} \ |
\ i = 1,2, 3 \}$.


\begin{theorem}
(Chru\'sciel and Galloway \cite{CG98})

There exists a connected set $K \subset R^2 = \{t = 0\} \subset
R^{2,1}$, where $R^{2,1}$ is a $2 + 1$ dimensional Minkowski
space-time, with the following properties:
\begin{enumerate}
\item The boundary $\partial K = \bar{K} - {\rm int}\, K$ of $K$ is a connected,
compact, Lipschitz topological submanifold of $R^2$. 
$K$ is the complement of a compact set $R^2$.
\item There exists no open set $\Omega  \subset R^{2,1}$ such that 
$\Omega \cap  H^+(K) \cap \{0 < t < 1\}$ is a differentiable submanifold of $R^{2,1}$.
\end{enumerate}
\end{theorem}

\begin{proposition}
(Beem and Kr\'olak \cite{BK98})

Let $W$ be an open subset of the Cauchy horizon $H^+(S)$.
Then the following are equivalent:
\begin{enumerate}
\item $H^+(S)$ is differentiable on $W$.
\item  $H^+(S)$ is of class $C^r$ on $W$ for some $r \geq 1$.
\item  $H^+(S)$ has no endpoints on $W$.
\item  All points of $W$ have multiplicity one.
\end{enumerate}
\end{proposition}

Note that the four parts of Proposition 1 are logically equivalent
for an {\em open} set $W$, but that, in general, they are not
necessarily equivalent for sets which fail to be open.  Using the
equivalence of parts (1) and (3) of Proposition 1, it now follows that
near each endpoint of multiplicity one there must be points where the
horizon fails to be differentiable.  Hence, each neighborhood of an
endpoint of multiplicity one must contain endpoints of higher
multiplicity. This yields the following corollary.

\begin{corollary}
(\cite{BK98})

If $p$ is an endpoint of multiplicity one on a Cauchy horizon
$H^+(S)$, then each neighborhood $W(p)$ of $p$ on $H^+(S)$ contains
points where the horizon fails to be differentiable.  Hence, the set
of endpoints of multiplicity one is in the closure of the crease set.
\end{corollary}

\section{A generic densely nondifferentiable Cauchy horizon}

We shall construct a densely nondifferentiable Cauchy horizon in the
$3$-di\-men\-sion\-al Minkowski space-time $R^{2,1}$, but our construction
can be generalized in a natural way to higher dimensions. Let $\Sigma$
be the surface ${t = 0}$, and let $K$ be a compact, convex subset of
$\Sigma$.  Let $\partial K$ denote the boundary of $K$. Let $\rho (x,
R)$ and $D(x, R)$ be respectively a circle and a disc with center at
$x$ and radius $R$.

\begin{definition}
%
A circle $\rho (x, R)$ is internally tangent to the boundary $\partial
K$ of $K$ if the disc enclosed by $\rho$ is contained in $K$ and for
all $\epsilon$ the disc of radius $R + \epsilon$ and center $x$ is not
contained in $K$.
\end{definition}

Let $\rho (x, R)$ be internally tangent to $\partial K$; then the point
$(x, R) \in R^{2,1}$ belongs to the future Cauchy horizon $H^+(K)$ and
conversely, if a point $(x, R) \in R^{2,1}$ belongs to $H^+(K)$ then
the circle $\rho (x, R)$ is internally tangent to $\partial K$.  If
$\rho (x, R)$ is internally tangent in at least two points of
$\partial K$ then it follows from Proposition 1 that $H^+(K)$ is not
differentiable at the point $(x, R)$ and the point $(x, R)$ has
multiplicity at least two.

We shall first construct a continuous curve that is not differentiable
on any open subset.  Let us take a line segment $l_0$ and let us
consider an isosceles triangle with base $l_0$ and let $\alpha_0$ be
the angle at the base and let $l_1$ denote the broken line consisting
of two equal arms of the triangle.  In the next step we construct two
isosceles triangles with bases that are segments of the broken
line $l_1$ and we choose the angles $\alpha_1$ at the base equal $q
\times \alpha_0$ where $q < 1/2$.  We iterate the above construction.
At the $N$th step of the construction the number of nondifferentiable
points of the curve increases by $2 N - 1$.  After the $N$th step of
the iterative procedure the vertex angle of the isosceles triangle
obtained in the $i$th step is given by
\be
\angle_N(x_i) = \pi - 
2 \alpha_1 \left [q^{i-1} - \frac{q^i - q^N}{1 - q}\right].
\ee
In the limit $N \rightarrow \infty$ the $i$th vertex angle is given by
$\pi - 2 \alpha_1 q^i \frac{q^{-1} - 2}{1 - q}$ 
and is strictly less than $\pi$ as $q < 1/2$.

Let us call the nowhere differentiable continuous curve constructed
above a {\em rough curve}.  Let us call a region of $\Sigma$ that is
bounded by a rough curve and two straight lines perpendicular to the
rough curve at its two endpoints\footnote{This notion is unambiguous,
as the slope of the rough curve at an endpoint is given by a
well-defined limit.} a {\em fan}.  The above construction
can be generalized to higher dimensions, for example in the
4-dimensional Minkowski space-time we construct a {\em rough surface}
in the following way. We consider a triangle and the first step is to
construct a pyramid with the triangle as a base and all angles between
the base and the sides of the pyramid equal to the same angle
$\alpha_1$; we then iterate the construction decreasing at each step
the angle $\alpha$ between the base and the sides of the pyramid by a
factor $q < 1/2$ as in the 3-dimensional case.  As a result we obtain
a nowhere differentiable surface and we define a 3-dimensional fan as
the region of $\Sigma$ bounded by the rough surface and planes
perpendicular to the rough surface passing through the sides of the
initial triangle.

\begin{theorem}
%
Let $b$ be a rough curve and $F$ the corresponding fan. 
Then the set of points of $F$ that are centers of circles tangent to $b$ 
in at least two points of $b$ is dense in the interior of the fan $F$.
\end{theorem}

\noindent 
{\em Proof:}

Each point of $F$ is the center of a circle tangent to $b$ at at least
one point.  If the claim of the theorem were false, then there would
exist a disc $D(x,R)$ with nonempty interior with the property that
every point $a \in {\rm int}\, D$ is the center of a circle tangent to the
rough curve at exactly one point.
\begin{enumerate}
\item  A vertex point cannot be a point of tangency of any circle with center
in ${\rm int}\,F$.
\item By construction the set of vertices of $b$ is dense in $b$.
Thus the complement of the set of vertices in $b$ is totally disconnected
(i.e. only one-element subsets are connected).
\end{enumerate}

Let us consider a map $P$ from the disc to $b$ that assigns to every
point $y$ of $D$ a point on $b$ that is tangent to the circle centered
at $y$. By assumption this point is unique and thus the map is
well-defined.

Let us show that the map $P$ is continuous. It is enough to prove that
if $a_n \rightarrow a$ then $P(a_n) \rightarrow P(a)$.  As $b$ is
compact, $P(a_n)$ has a subsequence that converges to a point $c$ on
$b$. Since the distance $d(a_n,P(a_n))$ is continuous on D we have
$d(c,a) = d(a,P(a))$.  Hence $c$ is a tangency point of a circle
centered at $a$ and consequently $c = P(a)$.

By the Darboux theorem the image $P(D(x,R))$ is connected and by 1. and
2. above, it is a one-point set. It then follows that R = 0 which is a
contradiction. {\bf QED}
\bigskip

The above theorem generalizes to the 3-dimensional case. In the case
of a 3-dimensional fan $F$ there exists a dense subset of $F$ such
that every ball with the center in this subset has at least two
tangency points to the rough surface. All steps of the proof of
Theorem 2 carry over to this case in the natural way.

Let ${\cal H}$ be the set of Cauchy horizons arising from compact
convex sets $K \subset \Sigma$ .  The topology on ${\cal H}$ is
induced by the Hausdorff distance on the set of compact and convex
regions K.

\begin{theorem}
%
Let ${\cal H}$ be the set of future Cauchy horizons $H^+(K)$ 
where $K$ are compact and convex regions of $\Sigma$.
The subset of densely nondifferentiable horizons is dense in ${\cal H}$.
\end{theorem}

\noindent
{\em Proof:}

Any compact and convex region K can be approximated in the sense of
Hausdorff distance by a (sequence of) convex polygons contained in
$K$.  Each of the vertex angles of such a polygon is strictly less
than $\pi$.  Over each side of the polygon we constract a rough curve
in such a way that the fans corresponding to the rough curves cover
the polygon.  This is always possible, since we may choose the
starting angle $\alpha_1$ in the rough curve's construction to obey
the condition
\be
\phi + \frac{2 \alpha_1}{1 - q} < \pi,
\ee
where $\phi$ is the largest vertex angle of the original polygon.
When $\alpha_1$ decreases to 0 the rough-edged polygon converges to
the original polygon in the sense of Hausdorff topology. {\bf QED}
\bigskip

It is clear that the above theorem generalizes to higher dimensions. 

\section{Some examples of densely nondifferentiable horizons}

In this Section we show that the construction of the previous Section
implies the existence of densely nondifferentiable Cauchy horizons of
partial Cauchy surfaces and also the existence of black hole event
horizons.

\begin{definition}
%
A {\em partial Cauchy surface} $S$ is a connected, acausal, edgeless
$n-1$ dimensional submanifold of $(M,g)$.
\end{definition}

\noindent
{\bf Example 1:} {\em A rough wormhole.}
\medskip

Let $R^{3,1}$ be the 4-dimensional Minkowski space-time and let $K$ be
a compact subset of the surface $\{t = 0\}$ such that its Cauchy
horizon is nowhere differentiable in the sense of the construction
given in Section 3. We consider a space-time obtained by removing the
complement of the interior of the set $K$ in the surface ${t = 0}$
from the Minkowski space-time.  Let us consider the partial Cauchy
surface $S = \{t = -1\}$.  The future Cauchy horizon of $S$ is the
future Cauchy horizon of set $K - {\rm edge}(K)$, since ${\rm
edge}(K)$ has been removed from the space-time.  Thus the future
Cauchy horizon is nowhere differentiable and it is generated by
past-endless null geodesics.  The interior of the set $K$ can be
thought of as a ``wormhole'' that separates two ``worlds'', one in the
past of surface $\{t = 0\}$ and one in its future.
\bigskip

\noindent 
{\bf Example 2:} {\em A transient black hole.}
\medskip

Let $R^{3,1}$ be the 4-dimensional Minkowski space-time and let $K$ be
a compact subset of the surface $\{t = 0\}$ such that its {\em past}
Cauchy horizon is nowhere differentiable in the sense of the
construction given in Section 3. We consider a space-time obtained by
removing from Minkowski space-time the closure of the set $K$ in the
surface
${t = 0}$.  Let us consider the event horizon E := $\dot{J}^{-}({\cal
J}^{+})$.  The event horizon $E$ coincides with $H^-(K)-{\rm edge}(K)$ and
thus it is not empty and nowhere differentiable.
The event horizon disappears in the future of surface $\{t = 0\}$ 
and thus we can think of the black hole (i.e. the set
$B := R^{3,1} - J^{-}({\cal J}^{+})$) in the space-time as ``transient''.   
\bigskip

\section{Acknowledgments}

The authors would like to thank P.T. Chru\'sciel for many helpful
discussions and a careful reading of the manuscript. This work was
supported by the Polish Committee for Scientific Research through
grants 2~P03B~130~16 and  2~P03B~073~15.

\end{document}